\documentclass[manuscript]{emulateapj}

\begin{document}

\title{Dynamical Treatment of Virialization Heating in Galaxy Formation}

\author{Peng Wang and Tom Abel}

\affil{Kavli Institute for Particle Astrophysics and Cosmology \\  Stanford Linear Accelerator Center and Stanford Physics Department, Menlo Park, CA 94025} 
\affil{Kavli Institute for Theoretical Physics, University of California, Santa Barbara, CA 93106}
\email{pengwang, tabel@stanford.edu}

\begin{abstract}
In a hierarchical picture of galaxy formation virialization continually transforms gravitational potential energy into kinetic  energies of the baryonic and dark matter. For the gaseous component the kinetic, turbulent energy is transformed  eventually into internal thermal energy through shocks and viscous dissipation. Traditionally this virialization and shock heating has been assumed to occur instantaneously allowing an estimate of the gas temperature to be derived from the the virial temperature defined from the embedding dark matter halo velocity dispersion. As the mass grows the virial temperature of a halo grows. Mass accretion hence can be translated into a heating term. We derive this heating rate from the extended Press Schechter formalism and demonstrate its usefulness in semi-analytical models of galaxy formation. Our method explicitly conserves energy unlike the previous impulsive heating assumptions. Our formalism can trivially be applied in all current semi-analytical models as the heating term can be computed directly from the underlying merger trees. Our analytic results for the first cooling halos and the transition from cold to hot accretion are in agreement with numerical simulations.
\end{abstract}

\maketitle

\section{Introduction}

In the traditional hierarchical scenario of galaxy formation \citep{rees_ostriker, white1, white2, kauffmann, cole}, it is assumed that the infalling gas is shock heated to the virial temperature of the hosting dark matter halo near the virial radius. However, recently, analytical argument \citep{dekel1}, cosmological simulations \citep{keres} and observations \citep{blanton2, kauffmann3, kauffmann2, milos1, milos2} have all shown evidence that  in halos below a critial mass scale $M_{cr}\sim 10^{12}$ M$_\odot$, infalling gas was not shock heated at the virial radius. Thus in addition to the traditional ``hot mode" of accretion, halos below the critical mass scale also have a ``cold mode" of accretion. It has been shown that introducing such a critical mass scale is advantageous in semi-analytical galaxy formation model \citep{dekel3, milos1, milos2}.

As the mass of a halo increases due to merging and accretion, the
temperature of the gas inside the halo will also change due to
turbulence and shocks which constantly transform the merging and
accreting kinetic energy into thermal energy. Since the
specific kinetic energy gained by halo during merging and accretion
is given by the change of virial temperature, the gas heating rate
can be computed as $\Gamma=k_B{dT_{vir}/dt}/(\gamma-1)$, where $\gamma$ is the adiabatic index which we will take to be $5/3$ as is the case for primordial gas. Since $T_{vir}\propto M^{2/3}$, this implies that the virialization heating rate is determined by the mass accretion history of the halo. This point was first proposed by \citet{yoshida} in the context of first structure formation.

Our basic assumption is that due to the turbulent nature of the virialization heating process, it is a local process acting on most gas particles inside the halo. This is very different from the popular picture that gas is virial heated by a spherical virial shock sitting roughly at the virial radius. The validity of this assumption can be seen from Fig. 5 of Wise \& Abel (2007), which shows two-dimensional slices of velocity divergence for four cosmological simulations with and without cooling. Velocity divergence is a good indicator of virialization shock positions. As can be seen in that figure, in the adiabatic cases, these shocks mainly exist at large radii where the gas virializes. But when radiative cooling is included, the turbulence becomes highly supersonic which then increases the frequency of shock fronts in the interior of the halo. Even gas deep inside the halo are affected by the cascading network of shocks in supersonic turbulence. This supports our basic assumption that the potential energy is first transformed to kinetic energy and is approximately shared with all the fluid in a dark matter halo. Under this assumption, we compute the average heating rate during accretion and merging. So whether the infalling gas can be shock heated at the virial radius depends on the competition between virialization heating rate and the local cooling rate: infalling gas can be shock heated at the virial radius if virialization heating rate is larger than the local cooling rate, otherwise infalling gas must flow cold to the inner halo.

In this work, we will compute this virialization heating rate and compare it to the gas cooling to find the evolution of the critical halo mass separating the cold and hot flow region in galaxy formation. We argue that our approach is advantageous as it formulates galaxy formation in an energy conserving fashion.

\section{Formalism}

In the spherical collapse model of halo formation \citep{gunn}, a halo formed at redshift $z$ with mass $M$
has a radius $R_{vir}$ that encloses a characteristic
overdensity of $\Delta_c$, which in a $\Lambda$CDM cosmology we will
take to be $178$ when $z\ge1$ and $356/(1+z)$ when $z<1$ (a more
precise fit is given by \citet{bryan}). So $M$ is related to $R_{vir}$ by 
$M=\Delta_c{4\pi\over3}R_{vir}^3\rho_m$, where $\rho_m(z)=(3H_0^2/8\pi G)\Omega_{m}(1+z)^3$ is the mean matter density at redshift $z$. For each halo, we define a virial velocity $V_{vir}^2={GM\over R_{vir}}$ and a virial temperature $T_{vir}={\mu m_p\over 2k_B}V_{vir}^2$ where $\mu$ is the mean molecular weight. From these definitions we find that
\begin{eqnarray}
T_{vir}&=&{\mu
m_pG^{2/3}\Delta_c^{1/3}\Omega_{m}^{1/3}H_0^{2/3}\over16^{1/3}k_B}M^{2/3}(1+z)\cr
&=& 4.8\times10^{-3}\left({M\over
M_\odot}\right)^{2/3}(1+z)\cr
&&\times\left({\Omega_{m}\over0.3}\right)^{1/3}\left({\Delta_c\over178}\right)^{1/3}\left({\mu\over0.59}\right)
{\rm K}\ ,\label{2.4}
\end{eqnarray}

 Using
${dz/dt}=-H_0[\Omega_{m}(1+z)^5+\Omega_{\Lambda}(1+z)^2]^{1/2}$ in
a $\Lambda$CDM cosmology and differentiating Eq. ({\ref{2.4}), one finds
\begin{eqnarray}
\Gamma&=&-{3\over2}{\mu
m_pG^{2/3}\Delta_c^{1/3}\Omega_{m}^{1/3}H_0^{5/3}\over54^{1/3}}M^{-1/3}{dM\over
dz}\cr
 && \times [\Omega_{m}(1+z)^7+\Omega_{\Lambda}(1+z)^4]^{1/2}  \cr &=&-2.95\times10^{-8}\left({M\over
M_\odot}\right)^{-{1\over3}}{d(M/M_\odot)\over
dz}\cr &&\times[\Omega_{m}(1+z)^7+\Omega_{\Lambda}(1+z)^4]^{1/2} \cr &&\times
\left({\Omega_{m}\over0.3}\right)^{{1\over3}}\left({\Delta_c\over178}\right)^{{1\over3}}\left({\mu\over0.59}\right)
{\rm eV}{\rm Gyr}^{-1} \ , \label{2.5}
\end{eqnarray}
where $dM/dz$ is the halo mass accretion rate, which
in simulations can be found routinely by constructing halo merger
trees. In this work, we will use a semi-analytical
approaches to compute $dM/dz$ proposed by \citet{bosch}. \citet{wechsler} have proposed a different form
of fitting formula for $dM/dz$, but their formula fitted well with
van den Bosch's formula with redefined parameters \citep{bosch}. 
Fig.~\ref{1} shows the mass accretion history of halos with current mass $10^{10}, 10^{11}, 2\times10^{12}, 10^{13}, 10^{14}, 10^{15}$ M$_\odot$  using van den Bosch's formula. 


\section{Implications}

\subsection{Virialization heating versus cooling}

To to see whether infalling gas can be shock heated to the virial temperature at the virial radius, we need to compare the heating rate (\ref{2.5}) to the gas cooling rate at the virial radius. 

To compute the cooling rate, when $T>10^4$ K, we use the tabulated cooling function for gas in collisional ionization equilibrium computed by \citet{sutherland}; when the gas temperature is smaller than $10^4$ K, atomic line cooling is insufficient and cooling rate is dominated by $H_2$ ro-vibriational transition, we will use the fitting formula for H$_2$ cooling function in \citet{ripamonti} assuming a universal H$_2$ fraction $f_{H_2}=10^{-3}$ \citep{tegmark, abel98}. 

Fig.\ref{2} shows the heating rate $\Gamma$ and cooling rate per particle $n(r_{vir})\Lambda(T_{vir})$ at virial radius for halos with current mass
$10^{10}, 10^{11}, 2\times10^{12}, 10^{13}, 10^{14}, 10^{15}$
M$_\odot$. We assume that the gas has an isothermal density profile and we plot the case for both metal-free and solar-metallicity gas. As can be seen from Fig. \ref{2}, in the case of metal-free gas, for $10^{10} (10^{11})$ M$_\odot$ halo, cooling
is always at least three (two) orders of magnitude larger than heating; for $2\times10^{12}$ M$_\odot$ halo,
cooling always dominates over heating but
they are comparable around $z\sim1$; for $10^{13}$ ($10^{14}, 10^{15}$) M$_\odot$ halo, heating overtakes cooling at
$z\sim3.5 (5, 6)$. In summary, in the case of metal free gas, for haloes with current mass smaller
than $2\times10^{12}$ M$_\odot$, cooling always dominate over heating, while for
halos with current mass larger than $2\times10^{12}$ M$_\odot$, heating will
overtake cooling when redshift is smaller than some critical
redshift $z_{cr}(M)$ which is a increasing function of current halo
mass.  In all cases, the heating rate decreases rapidly when $z\rightarrow 0$ due to the rapidly decreasing accretion rate in a dark energy dominated Universe. 

From Fig. \ref{2} we can see that cooling dominates heating around redshift $10$ while
the cooling rate is still dominated by atomic line cooling. So our framework predicts that gas falling into halo at redshift $\sim10$ generally cannot be shock heated at the edge of the halo if their virial temperature are greater than $10^4$ K. 

Abel et al. (2000, 2002) showed that the first generation of stars formed inside halos of mass $\sim 10^6$ M$_\odot$ at redshift $>20$. From Fig. \ref{1}, we can see that this corresponds roughly to halo of current mass $10^{10}$ M$_\odot$. Then from Fig. \ref{2}, we can see that virialization heating is comparable to the cooling rate of those minihalos at high redshift, consistent with the conclusion of \citet{yoshida}.

\subsection{Critical mass scale for shock heating at small redshift}

As we have seen in Fig. \ref{2}, virialization heating dominates cooling at small redshift for halos above a ciritcal mass. Fig. \ref{3} shows the evolution of the critical halo mass $M_{cr}$ where the rates are equal. We have considered both an isothermal density profile and an NFW profile \citep{nfw}. For NFW profile, we took the concentration parameter $c=12$, so the density at the virial radius is $0.17$ of the mean density of the halo. It can be seen that for zero metallicity, $M_{cr}\sim 10^{11.5}-10^{12}$ M$_\odot$, while for solar metallicity, $M_{cr}\sim 10^{12.5}-10^{13}$, with both of them changing very slowly at redshift range $0.5<z<3$. This is consistent with recent cosmological simulations which found that below redshift
$3$, only gas in halos above an almost redshift independent critical mass $M_{cr}\sim10^{12}$ M$_\odot$
 can be shock heated at the virial radius while below this mass there is a cold mode of gas accretion \citep{keres, dekel1, dekel2, dekel3}. Using a spherical symmetric stability analysis, \citet{dekel1}
explained this to be due to the instability of virial shock in small
halos. Our result implies a complementary
interpretation of this phenomenon. Note that $M_{cr}$ increases rapidly when $z<0.5$ because halo accretion rate decrease rapidly due to the onset of dark energy domination. 

\subsection{Implications for semi-analytical galaxy formation models}

A central task of semi-analytical galaxy formation models (SAM) is computing the evolution of hot and cold gas fractions inside dark matter halos \citep{white1, white2, kauffmann, cole, springel}. In traditional SAM, cooling is treated explicitly as a dynamical process while heating is treated impulsively. In SAM cooling is the result of competition between the local cooling time and dynamical time. The point where cooling time equals dynamical time is defined to be the ``cooling radius" and is used to compute the amount of cold gas and then star formation rate. For heating, traditional SAM just assumes that when halo merge with each other, all gas that is not already cooled is shock-heated to the virial temperature of the new halo \citep{kauffmann, cole}. We will see that if we treat heating also as a dynamical process, this is usually not the case. 

Firstly, we define a heating radius $r_{h}$ as $n(r_{h})\Lambda(T_{vir})=\Gamma$. So gas can be shocked to the new virial temperature during merger for gas lying between the heating radius $r_{h}$ and $R_{vir}$. Following SAM \citep{kauffmann, cole}, consider an isothermal gas density profile
$n_g(r)=M_g/(4\pi \mu m_p R_{vir}r^2)$ where $M_g$ is the total gas mass, $\mu$ the mean molecular weight, then $r_{h}$ becomes
\begin{equation}
r_{h}=\left(\frac{M_g\Lambda(T_{vir})}{4\pi \mu m_pR_{vir}\Gamma}\right)^{1/2} \ . \label{rrh}
\end{equation}

In traditional SAM, it is assumed that all gas between the cooling radius and $R_{vir}$ can be heated to the virial temperature of the new halo in mergers. Fig. \ref{4} shows the evolution of the heating radius (\ref{rrh}) and the cooling radius defined by \citet{white2} for halos of current mass $ 10^{12}, 10^{13}, 10^{14}, 10^{15}$ M$_\odot$. From Fig. \ref{4} we can see that the heating radius is always at least two times larger than the cooling radius for both the zero and solar metallicity cases. Therefore, the typical mass accretion rate does not supply sufficient energy to heat all of the gas at $r>r_{cool}$. \emph{Hence traditional SAM may significantly overestimated the amount of gas that can be shock heated during galaxy mergers}. This has direct implications for using SAM to compute thermal radiation, cosmic ray acceleration, AGN feedback, or whatever process that depends on the amount of halo hot gas. The kinetic energy associated to the gas accreted in code mode just lost as cooling radiations. The gas that has been shocked heated at one redshift may not be shocked heated to the new virial temperature at a later merger. Their temperature will keep to be the virial temperature corresponding to the time they cross the heating radius. However, note that this is true only when that gas is still outside the cooling radius, otherwise it will just cool down to the minimum temperature allowed by the gas cooling properties. Thus, instead of assuming all uncooled gas to have the same virial temperature of the new halo, a more consist way is to compute a temperature profile using the virialization heating rate if what is to be calculated depends on the amount of hot gas.

Using $n_g^2(r)\Lambda(T)-n_g(r)\Gamma$ as the net cooling rate, one can estimate the change in $r_{cool}$. Thus the local cooling time is given by
\begin{equation}
t_{cool}={3k_BT_{vir}n_g(r)\over2 [n_g^2(r)\Lambda(T_{vir})-n_g(r)\Gamma]} \ ,\label{2.7}
\end{equation}
where $n_g(r)$ is the isothermal gas density profile.

If one assumes the density profile remains to be approximately fixed during cooling, the gas in the halo will have cooled at time $t$ out to a radius $r_{cool}$ determined by the equality of cooling time and halo dynamical time \citep{white1}
\begin{equation}
t_{cool}(r_{cool}(t))=t=t_{dyn}={R_{vir}\over V_{vir}} \ .\label{2.8}
\end{equation}

Hence the heating corrected cooling radius is given by
\begin{equation}
r_{cool}=\left[\frac{M_g\Lambda(T_{vir})}{4\pi\mu m_p[{3\over2}k_BT_{vir}V_{vir}+R_{vir}\Gamma]}\right]^{1/2} \ .\label{2.9}
\end{equation}
When $\Gamma=0$, this reduces to the usual formula of cooling radius \citep{white2}, as expected.

Fig. \ref{4} shows the evolution of cooling radius with and without the heating correction term (\ref{2.9}). It can be seen that the amout of cooled gas is decreased only by a small amout compared to the case without heating correction. So virialization heating, while important for determining whether gas can be shock heated, plays a lesser role in influencing gas cooling inside halos. This explains why the amount of cold gas predicted by SAM and numerical simulation compares well \citep{yoshida2, helly}.

From Eqs. (\ref{2.7}) and (\ref{2.8}), we have
\begin{eqnarray}
{dr_{cool}\over dt}&=&{r_{cool}\over 2 t_{cool}}{n_g(r_{cool})\Lambda(T_{vir})-\Gamma\over n_g(r_{cool})\Lambda(T_{vir})}\label{2.11}
\end{eqnarray}

So combining Eqs. (\ref{2.9}) and (\ref{2.11}), the cooling rate $\dot{M}_{cool}={dM_{cool}/ dt}=4\pi\rho_g(r_{cool})r^2_{cool}{dr_{cool}/dt}$ with heating correction is given by
\begin{equation}
\dot{M}_{cool}={M_gV_{vir}r_{cool}\over2R_{vir}^2}{n_g(r_{cool})\Lambda(T_{vir})-\Gamma\over n_g(r_{cool})\Lambda(T_{vir})}\label{2.12}
\end{equation}
where $r_{cool}$ is given by Eq. (\ref{2.9}).

In SAM, $\dot M_{cool}$ is a key quantity to compute the evolution of cold gas and thus star formation rate \citep{cole}. Eq. (\ref{2.12}) shows explicitly how this quantity will be modified in our formalism and thus can be directly applied to SAM. Furthermore, although we assumed $\Gamma$ contains only virialization heating in this work, the final result Eq. (\ref{2.12}) is actually general which shows how to consistently incorporate heating effect into SAM. For example, it can also be applied to discuss AGN heating (see e.g. Croton et al. 2006).

In summary, we have found that gas heating is determined by the competition between cooling rate and heating rate while gas cooling is determined by the competition between cooling time and dynamical time. Our analysis suggests that just like current treatment of gas cooling in SAM, we should also treat gas heating as a dynamical and local process, especially when we try to compute physical quantities that relies on the amount of hot gas inside halos, e.g. thermal radiation from halo hot gas (e.g. Miniati et al. 2004, Furlanetto et al. 2005).

\section{Conclusions and discussions}

We have computed a virialization heating rate which is directly related to the halo mass accretion history using van den Bosch's fitting formula.

By comparing the virialization heating rate to the cooling rate, we find that gas can be shocked heated at the virial radius only for large halos at low redshift and small halos at very high redshift. The critical halo mass computed in our framework agrees with recent simulations and other analytical arguments.

Using the virialization heating rate, we also found that current SAMs may have significantly overestimated the amount of gas that can be shock heated.  Our formalism provides an energy conserving remedy to this problem. On the other hand, gas cooling is primarily determined by the competition between cooling time and dynamical time, which explains the good fit of the cold gas amount in the literature by comparing SAMs to numerical simulation.

Due to the energy-conserving nature of our formalism, it is also suitable to be used to compute quantities such as the thermal radiation from halo hot gas. Furthermore, since cosmic ray acceleration and generation of galactic magnetic field are directly related to the amount of shocked gas \citep{loeb1, loeb2, keshet, medvedev, fields}, our formalism can also be applied to compute these processes.

Finally, we would like to indicate that an important issue for our calculation is the absence of scatter in van den Bosch's formula as it is the averaged mass accretion rate in the sense of both space and time. As a space average, it implies halos corresponding to high sigma peaks will have larger mass accretion rate than van den Bosch's formula and vice versa for halos corresponding to low sigma peaks. As a time average, it implies that for a single halo, accretion rate can become larger than van den Bosch's formula during major merger and smaller in the quiescent accretion epochs. Realizing the time-average nature of van den Bosch's formula may be quite important, since the final state of a halo may be quite different if one computes the heating rate using the true mass accretion rate constructed from simulation rather than van den Bosch's formula. However, we also note that from Figures 2 and 3 in \citet{bosch}, it's not a bad estimate that the scatter is roughly $0.5M(z)$, almost independent of redshift. From Eq. (\ref{2.5}), we expect that this leads to a scatter of $\sim 1.5^{2/3}\approx 1.3$ in the heating rate. Our calculation captures some of the essential physical effects of the virialization heating process and is easy to implement in SAMs that use merger trees derived from N-body simulations or from Monte-Carlo techniques.

\section*{Acknowledgements}
This work was supported by NSF CAREER award AST-0239709 from the National Science Foundation. We would like to thank Naoki Yoshida for helpful comments on the draft.

\begin{figure*}
   \includegraphics[height=.4\textheight]{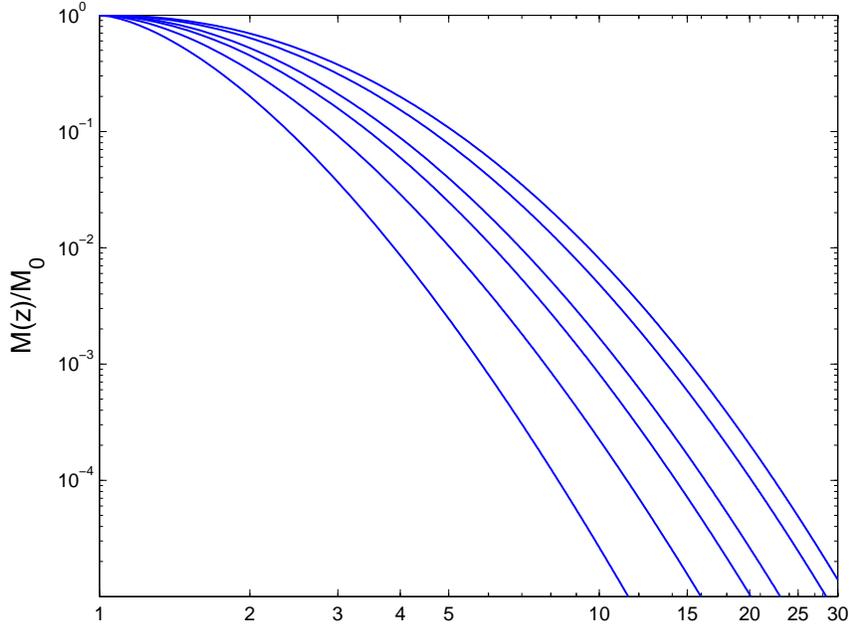}
    \caption{van den Bosch's fitting of halo mass accretion history for halos with current masses $10^{10}, 10^{11}, 10^{12.3}, 10^{13}, 10^{14}, 10^{15}$ M$_\odot$ from top to bottom.}\label{1}
\end{figure*}

\begin{figure*}
   \includegraphics[height=.4\textheight]{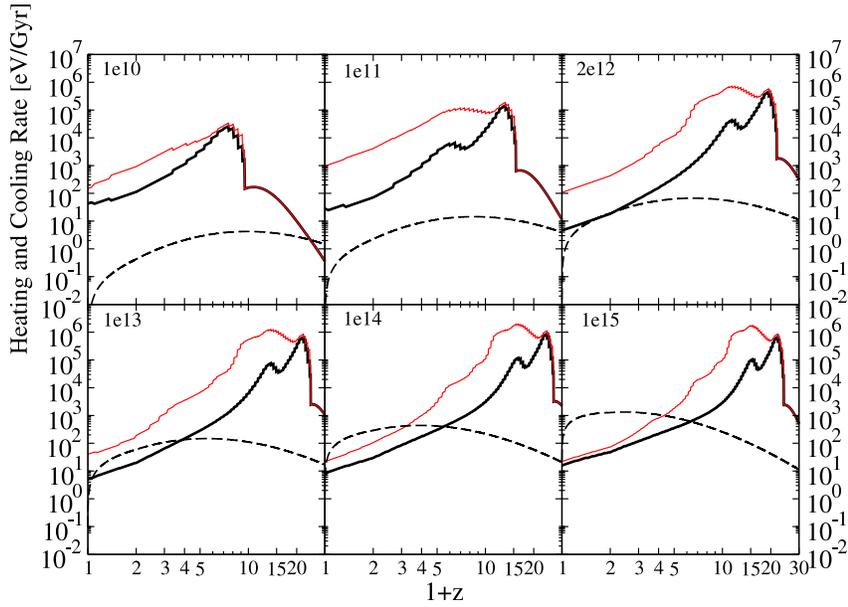}
    \caption{Heating and cooling rate versus redshift. Dashed lines are heating rate for halo of current mass
$10^{10}, 10^{11}, 2\times10^{12}, 10^{13}, 10^{14}, 10^{15}$ M$_\odot$. Thick solid lines are the cooling rate at the virial radius assuming zero metallicity. Thin solid lines are the cooling rate at the virial radius assuming solar metallicity.}\label{2}
\end{figure*}

\begin{figure}
  \includegraphics[height=.4\textheight]{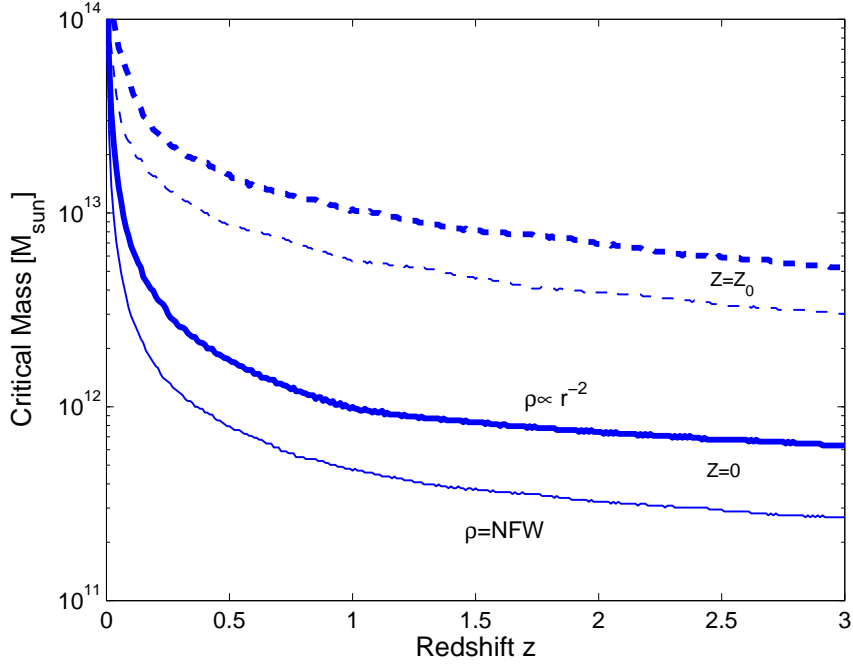}
    \caption{Evolution of the critical halo mass scale above which the heating rate is larger than cooling rate at the virial radius. Solid lines are computed assuming zero-metallicity while dashed lines are solar metallicity. Thick lines are computed assuming isothermal density profile while thin lines are NFW profile with concentration $c=12$.}\label{3}
\end{figure}

\begin{figure}
  \includegraphics[height=.4\textheight]{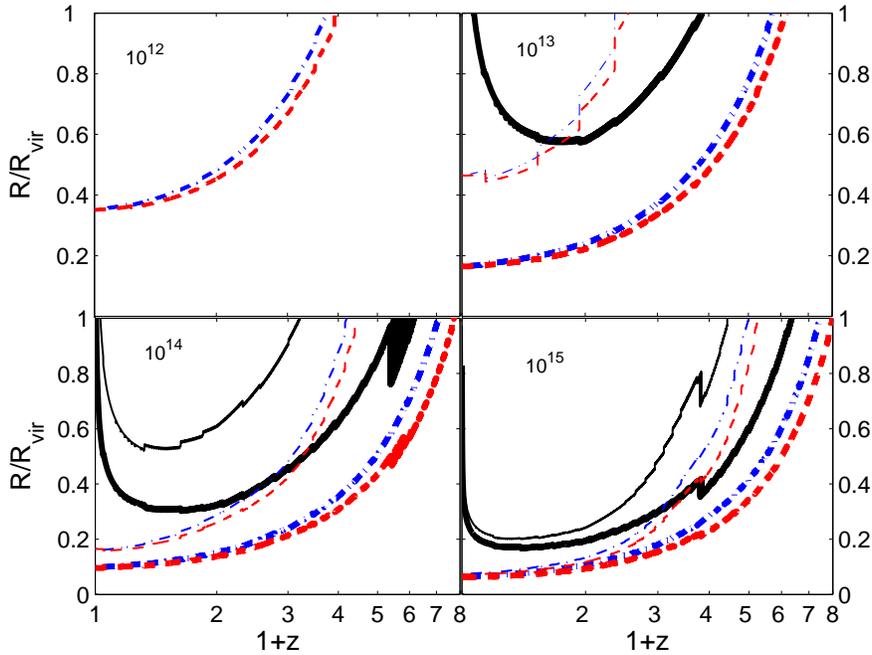}
    \caption{Evolution of heating radius (solid) and cooling radius with (dashed) and without (dot-dashed) heating correction in unit of the virial radius for halos with mass $10^{12}, 10^{13}, 10^{14}, 10^{15}$ M$_\odot$ today (beginning from the top-left plot and mass increases clockwise). Thick lines are for zero metallicity while thin lines are for solar metallicity. Note that the $10^{12}$ M$_\odot$ halo plot does not have a heating curve because in that case cooling dominates heating at the virial radius.} \label{4}
\end{figure}


\begin{thebibliography}{99}

\bibitem[Abel et al. (1998)]{abel98}
Abel, T.,  Anninos, P., Norman, M. L. \& Zhang, Y., 1998, \apj, 508, 518.

\bibitem[Abel et al. (2000)]{abn00}
Abel, T., Bryan, G. \& Norman, M., 2000, \apj, 540, 39.

\bibitem[Abel et al. (2002)]{abn02}
Abel, T., Bryan, G. \& Norman, M., 2002, Science, 295, 93. 


\bibitem[Birnboim \& Dekel (2003)]{dekel1}
Birnboim, Y. \& Dekel, A., 2003, \mnras, 345, 349.


\bibitem[Blanton et al. (2005)]{blanton2}
Blanton, M. R., Eisentstein, D., Hogg, D. W., Schlegel, D. J. \& Brinkmann, J., 2005, \apj, 629, 143. 

\bibitem[Bryan \& Norman (1998)]{bryan}
Bryan, G.L. \& Norman, M.L., 1998, \apj, 495, 80.

\bibitem[Cattaneo et al. (2006)]{dekel3}
Cattaneo, A., Dekel, A., Devriendt, J., Guiderdoni, B. \& Blaizot, J.,
astro-ph/0601295.

\bibitem[Ciotti \& Ostriker (2001)]{ostriker}
Ciotti, L. \& Ostriker, J.P., 2001, \apj, 551, 131.

\bibitem[Cole et al. (2000)]{cole}
Cole, S., Lacey, C. G., Baugh, C. M. \& Frenk, C. S., 2000, \mnras, 319, 168.

\bibitem[Cooray \& Milosavljevic (2005a)]{milos1}
Cooray, A. \& Milosavljevic, M., 2005a, \apj, 627, L85.

\bibitem[Cooray \& Milosavljevic (2005b)]{milos2}
Cooray, A. \& Milosavljevic, M., 2005b, \apj, 627, L89.

\bibitem[Croton et al.(2006)]{croton} 
Croton, D.~J., et al.\ 2006, \mnras, 365, 11 

\bibitem[Dekel \& Birnboim (2004)]{dekel2}
Dekel, A. \& Birnboim, Y., 2004, astro-ph/0412300.

\bibitem[Furlanetto et al. (2005)]{furlanetto05}
Furlanetto, S.~R., Schaye, J., Springel, V. \& Hernquist, L., 2005, \apj, 622, 7. 

\bibitem[Gunn \& Gott (1972)]{gunn}
Gunn, J. E. \& Gott, J. R., III, 1972, \apj, 176 1.

\bibitem[Helly et al. (2003)]{helly}
Helly, J. C., Cole, S., Frenk, C. S., Baugh, C. M., Benson, A., Lacey, C. \& Pearce, F. R., 2003, \mnras, 338, 913.

\bibitem[Hernquist \& Springel (2003)]{springel}
Hernquist, L. \& Springel, V., 2003, \mnras, 341, 1253.

\bibitem[Kauffmann et al. (1999)]{kauffmann}
Kauffmann, G., Colberg, J. M., Diaferio, A. \& White, S. D. M., 1999, \mnras, 303, 188.

\bibitem[Kauffmann et al. (2003a)]{kauffmann3}
Kauffmann, G., et al., 2003, \mnras, 341, 54.

\bibitem[Kauffmann et al. (2003b)]{kauffmann2}
Kauffmann, G., et al., 2003, \mnras, 341, 33.

\bibitem[Keres et al. (2005)]{keres}
Keres, D., Katz, N., Weinberg, D. H. \& Dave, R., 2005, \mnras, 363, 2.

\bibitem[Keshet et al. (2003)]{keshet}
Keshet, U., Waxman, E., Loeb, A., Springel, V. \& Hernquist, L.,
2003, \apj, 585, 128.




\bibitem[Loeb \& Waxman (2000)]{loeb2}
Loeb, A. \& Waxman, E., 2000, Nature, 405, 156.

\bibitem[Medvedev et al. (2005)]{medvedev}
Medvedev, M. V., Silva, L. O. \& Kamionkowski, M., astro-ph/0512079.

\bibitem[Miniati et al. (2004)]{miniati}
Miniati, F., Ferrara, A., White, S. D. M. \& Bianchi, S., 2004, \mnras, 348, 946.

\bibitem[Navarro et al. (1997)]{nfw}
Navarro, J. F., Frenk, C. S. \& White, S. D. M., 1997, \apj, 490, 493.

\bibitem[Pavlidou \& Fields (2006)]{fields}
Pavlidou, V. \& Fields, B. D., 2006, astro-ph/0611923.

\bibitem[Press \& Schechter (1974)]{ps}
Press, W. H. \& Schechter, P., 1974, \apj, 187, 425.

\bibitem[Rees \& Ostriker (1978)]{rees_ostriker}
Rees, M. \& Ostriker, J. P., 1978, \mnras, 179, 541. 

\bibitem[Ripamonti \& Abel (2005)]{ripamonti}
Ripamonti, E. \& Abel, T., 2005, astro-ph/0507130.

\bibitem[Sheth \& Tormen (1999)]{sheth}
Sheth, R. K. \& Tormen, G., 1999, \mnras, 308, 119.

\bibitem[Sutherland \& Dopita (1993)]{sutherland}
Sutherland, R. S. \& Dopita, M. A., 1993, \apjs, 88, 253.

\bibitem[Tegmark et al. (1997)]{tegmark}
Tegmark, M., Silk, J., Rees, M. J., Blanchard, A., Abel, T. \& Palla, F., 1997, \apj, 474, 1.


\bibitem[van den Bosch (2002)]{bosch}
van den Bosch, F. C., 2002, \mnras, 331, 98.

\bibitem[Waxman \& Loeb (2000)]{loeb1}
Waxman, E. \& Loeb, A., 2000, \apj, 545, L11.

\bibitem[Wechsler et al. (2002)]{wechsler}
Wechsler, R. H., Bullock, J. S., Primack, J. R., Kravtsov, A. V. \&
Dekel, A., 2002, \apj, 568, 52.

\bibitem[White \& Frenk (1991)]{white2}
White, S. D. M. \& Frenk, C. S., 1991, \apj, 379, 52.

\bibitem[White \& Rees (1978)]{white1}
White, S. D. M. \& Rees, M. J., 1978, \mnras, 183, 341.

\bibitem[Wise \& Abel (2007)]{wise}
Wise, J. \& Abel, T.\ \apj, accepted.

\bibitem[Yoshida et al. (2003)]{yoshida}
Yoshida, N., Abel, T., Hernquist, L. \& Sugiyama, N., 2003, \apj, 592, 645.

\bibitem[Yoshida et al. (2002)]{yoshida2}
Yoshida, N., Stoehr, F., Springel, V. \& White, S. D. M., 2002, \mnras, 335, 762. 

\end{thebibliography}
\end{document}